\documentclass[conference]{IEEEtran}
\IEEEoverridecommandlockouts
\usepackage{cite}
\usepackage{amsmath,amssymb,amsfonts}
\usepackage{algorithmic}
\usepackage{graphicx}
\usepackage{textcomp}
\usepackage{xcolor}
\usepackage{caption}
\usepackage{subcaption}

\def\BibTeX{{\rm B\kern-.05em{\sc i\kern-.025em b}\kern-.08em
    T\kern-.1667em\lower.7ex\hbox{E}\kern-.125emX}}

\usepackage{tikz}
\usepackage{textcomp}
\usepackage{hyperref}
\usepackage{lipsum}

\newcommand\copyrighttext{%
	\footnotesize \textcopyright 2023 IEEE. Personal use of this material is permitted.
	Permission from IEEE must be obtained for all other uses, in any current or future
	media, including reprinting/republishing this material for advertising or promotional
	purposes, creating new collective works, for resale or redistribution to servers or
	lists, or reuse of any copyrighted component of this work in other works.}
\newcommand\copyrightnotice{%
	\begin{tikzpicture}[remember picture,overlay]
		\node[anchor=south,yshift=10pt] at (current page.south) {\fbox{\parbox{\dimexpr\textwidth-\fboxsep-\fboxrule\relax}{\copyrighttext}}};
	\end{tikzpicture}%
}

\begin{document}

\title{Performance Evaluation of Transport Protocols and Roadmap to a High-Performance Transport Design for Immersive Applications  \\
\thanks{This work is supported by Electronics and Telecommunications Research Institute (ETRI) grant funded by ICT R\&D program of MSIT/IITP[2021-0-00715, Development of End-to-End Ultra-high Precision Network Technologies].}
}

\author{\IEEEauthorblockN{Inayat Ali}
\IEEEauthorblockA{\textit{} 
\textit{ETRI}\\
Daejeon, South Korea \\
inayat@etri.re.kr}
\and
\IEEEauthorblockN{Seungwoo Hong}
\IEEEauthorblockA{\textit{} 
\textit{ETRI}\\
Daejeon, South Korea \\
swhong@etri.re.kr}
\and
\IEEEauthorblockN{Pyung-koo Park}
\IEEEauthorblockA{\textit{} 
\textit{ETRI}\\
Daejeon, South Korea\\
parkpk@etri.re.kr}
\and
\IEEEauthorblockN{Tae Yeon Kim}
\IEEEauthorblockA{\textit{} 
\textit{ETRI}\\
Daejeon, South Korea \\
tykim@etri.re.kr}  
}

\maketitle
\copyrightnotice

\begin{abstract}
Immersive technologies such as virtual reality (VR), augmented reality (AR), and holograms will change users' digital experience. These immersive technologies have a multitude of applications, including telesurgeries, teleconferencing, Internet shopping, computer games, etc. Holographic-type communication (HTC) is a type of augmented reality media that provides an immersive experience to Internet users. However, HTC has different characteristics and network requirements, and the existing network architecture and transport protocols may not be able to cope with the stringent network requirements of HTC. Therefore, in this paper, we provide an in-depth and critical study of the transport protocols for HTC. We also discuss the characteristics and the network requirements for HTC. Based on the performance evaluation of the existing transport protocols, we propose a roadmap to design new high-performance transport protocols for immersive applications. 

\end{abstract}


\section{Introduction}
Virtual reality (VR), augmented reality (AR), mixed reality (MR), and holographic technologies are examples of immersive technologies. These technologies have gained much attention from industry and academia in recent years. Applications of immersive technologies include but are not limited to holographic teleportation, tactile robotic surgery, virtual orchestra, E-commerce, and video games. Holographic-type communication is an emerging type of communication that involves the transmission and interaction of volumetric holograms over the Internet. For example, holographic teleportation will facilitate real-time interaction with the remote participants of a meeting by projecting them into the meeting room in the form of a holographic object. Similarly, holographic telesurgery will allow surgeons to perform surgery with more ease, and clarity \cite{b3}. HTC and immersive applications are used interchangeably in this article since both are immersive technologies and generate volumetric data.
The need to handle this kind of immersive applications and the volumetric data generated by these has become vital as big Internet giants like Facebook \cite{b1} and Roblox kick-started their immersive services. Roblox's monthly active users are over $150$ million \cite{b2}; imagine if most services on their platform go immersive, this will put a heavy burden on network resources. 
Particularly, we focus on holographic-type communication (HTC). A plethora of Internet applications, e.g., telesurgery, virtual orchestra, and tele-conferecing, are on their way to adopting holographic technologies. Moreover, these applications will put significant demands on network infrastructures, such as high bandwidth, ultra-low latency, and synchronization of multiple data streams, which the current network standards, especially the transport protocols, do not support.

Novel network architecture and transport protocols are needed to combat the requirement of HTC. The current standard of transport protocols may not be able to handle this kind of volumetric data while maintaining other application-specific requirements such as reliability, application-specific latency, and on-time delivery of specific data. At the transport layer, video streaming services either use TCP for quality or UDP for low latency and high throughput. TCP-based solutions include HTTP with adaptive streaming (HAS), and real-time media protocol (RTMP)\cite{b4}. UDP-based solutions, which are adopted for low latency, include the real-time protocol (RTP) combined with RTCP. Novel approaches are mostly built on top of UDP, such as QUIC \cite{b5} and WebRTC \cite{b6}. For designing a novel transport layer solution for HTC, it is necessary to first deeply study the working and performance of current transport layer standards for HTC. Until now, these standard protocols have not been evaluated for holographic-type communication. QUIC, particularly is a new protocol and has never been critically studied in comparison with the existing protocol for HTC. 

The contribution of this paper is threefold:

 (1). This paper provides a detailed evaluation and comparison of TCP, UDP, and QUIC protocols for HTC.

 (2). Discuss the challenges in the existing transport protocol and network requirements of HTC and other immersive network applications.

 (3). Present a roadmap to the design of novel high-performance transport protocols for immersive network applications that will guide our future research.

The rest of the paper is organized as follows. We discuss the challenges and network requirements of holographic-type communication in section $2$. In section $3$, we give an overview of the standard transport protocols. We provide critical discussion and performance analysis of the UDP, TCP, and QUIC under different congestion variants in section $4$. In section $5$, a roadmap to designing novel transport protocols for HTC will be presented, and then we will conclude the work in section $6$.

\section{Challenges and Network Requirements of Immersive Applications}
The existing transport protocols have lesser to no communication with the application and network layer making it challenging to ensure application demands. Therefore, developers are considering application-level solutions to cope with the transport layer limitations, e.g. WebRTC and MPEGDASH. However, their application to HTC has yet to be tested. 

Holograms are volumetric objects and require high bandwidth even after applying high-quality compression techniques. HTC, therefore, have very stringent network requirements. HTC objects require continuous user interactivity, whether streamed in real-time or dealing with offline content, as the user will still interact with the object by changing the viewing angle and position. The portion of the HTC object visible from the current view angle must be retrieved first compared to the data corresponding to the portion not visible from the current angle. This user interactivity is difficult to achieve as predicting the current view angle is challenging. This nature of HTC content in immersive applications demands ultra-high bandwidth and ultra-low latency.

\subsection{High Bandwidth}

The bandwidth demand of HTC will increase by several orders of magnitude over 3D VR content. For example, a Microsoft Kinect for Windows v2 generates nearly $70$ Mbits of raw data per frame or $2$ Gbits of raw data per second at $30$ FPS. After applying high-quality compression techniques, this data can be reduced to approximately $40$ Mbits per second per single camera sensor. The quality of a hologram depends on the number of sensors used to capture a hologram. A minimum of $8$ sensors will generate over $300$ Mbits of data per second. The amount increases with the increased quality of images captured, the number of camera sensors, and the number of viewpoints. Hence very high bandwidth will be required to stream HTC.

\subsection{Ultra-Low Latency}
Ultra-low latency will be required for a truly immersive experience. However, ultra-low latency for such volumetric data with high bandwidth requirements is hard to achieve. The latency requirement becomes very challenging in the case of HTC. Because not only real-time HTC streaming requires ultra-low latency, but also static HTC content needs ultra-low latency as the user may change the view angle continuously, which will require the application to retrieve part of the HTC content visible from that view angle. So this continuous user interactivity with HTC objects makes the latency requirement of HTC more challenging. 

\subsection{Application Requirements}
Different network applications have different requirements from the network. Some applications are delay sensitive, and some are loss sensitive. These requirements are common in traditional network applications, but in immersive applications, along with in-time delivery, on-time delivery of data may also be vital, for example, in telesurgery. Data spread over multiple streams must be received on time to avoid inconsistency and ensure synchronization. These strict application requirements must be addressed while designing standards for holographic-type communication. 

\section{Transport Protocols Overview}
\begin{figure}[b]
	\centering
	\includegraphics[scale=0.5]{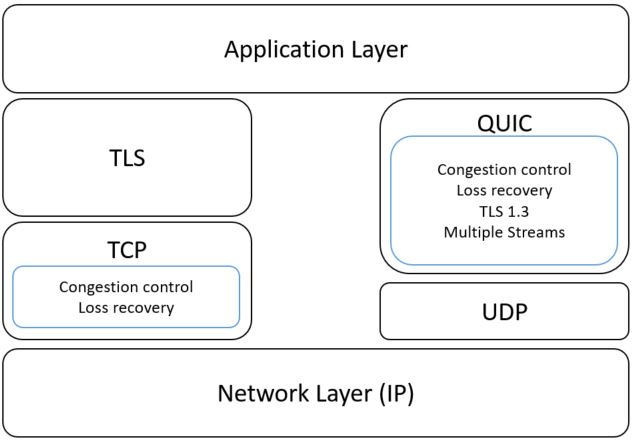}
	\caption{QUIC and TCP protocols stack}
	\label{fig:0.2}
\end{figure}

Until recently, only UDP and TCP were the dominant transport layer protocols. Different applications adopt either UDP or TCP based on the application requirements. Though fast, UDP is a connection-less protocol used by applications that do not require guaranteed delivery like most video streaming and broadcasting applications. Applications that require guaranteed delivery of data and quality use TCP. Many transport protocols, such as stream control transmission protocol (SCTP) \cite{b9}, were developed with time. However, due to the design of the Internet, these protocols could not be used on the public scale because the middleboxes usually drop packets other than UDP and TCP. QUIC \cite{b11}, developed by google and deployed in past years, is now becoming the most dominant transport protocol since it is integrated into google and YouTube platforms. QUIC runs on top of UDP, and most of the QUIC header is encrypted, enabling it to traverse the Internet without being dropped by the middleboxes. The QUIC and TCP stack has been shown in figure \ref{fig:0.2}.

QUIC overcomes many problems in TCP, e.g., Head of line blocking (HoL), long handshake delay, packet reordering issues, etc. A single QUIC connection is composed of multiple streams recognized by stream id. These streams can be mapped to HTTP/2 streams. This feature is very effective as the loss of packets in one stream does not block other streams in the same connection hence resolving the HoL issue. Furthermore, QUIC connection establishment requires one RTT compared to TCP/TLS $1.3$, which requires $2$ RTTs. QUIC also introduced $0$ RTT connection establishment for recently closed connections. QUIC also resolves the retransmission ambiguity in TCP, where the TCP can not recognize whether the ACK is for the first transmission or retransmission. QUIC uses per stream buffer at sender and receiver that ameliorates HoL issues among streams. 

\section{Performance Evaluation of Transport Protocols for Immersive Applications}

We perform a critical analysis of the standard transport protocols, including QUIC and congestion variants like YeAH \cite{b7} and  Vegas \cite{b8}. We use immersive applications like volumetric data in different network scenarios to check the performance of the transport protocols. We compare the performance of the standard transport protocols based on end-to-end delay, throughput, jitter, delivery ratio, and retrieval time of the HTC objects. We also analyze the protocols by looking at the RTT variations, congestion window, and queue occupancy. 

\subsection{Simulation Environment}
\begin{figure}[b]
	\centering
	\begin{subfigure}[b]{0.241\textwidth}
		\centering
		\includegraphics[width=\textwidth]{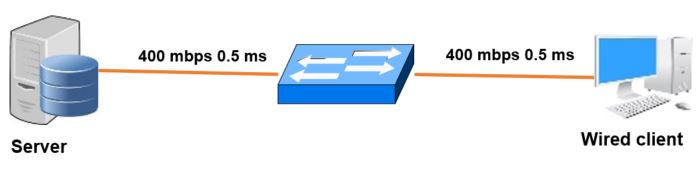}
		\caption{Simple LAN scenario}
		\label{fig:0of1}
	\end{subfigure}
	\hfill
	\begin{subfigure}[b]{0.241\textwidth}
		\centering
		\includegraphics[width=\textwidth]{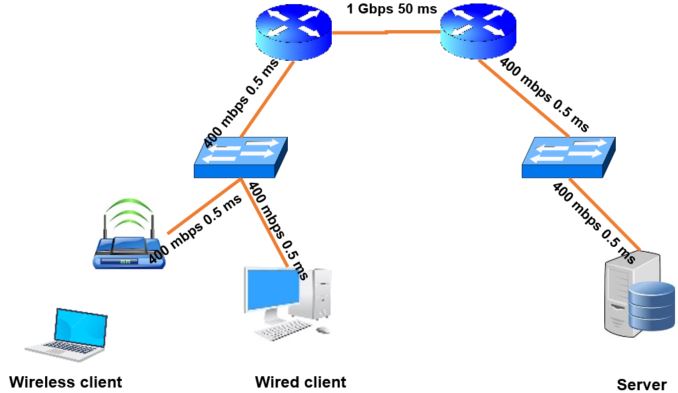}
		\caption{High latency WAN scenario}
		\label{fig:0of2}
	\end{subfigure}
	\caption{Network scenario for performance evaluation}
	\label{fig:0.1}
\end{figure}

We use ns-3, an open-source network simulator, for creating the simulation environment. We created two network scenarios, i.e., a simple LAN scenario to see the performance of the protocols at full throttle and a high latency WAN scenario to analyze the performance of transport protocols in a dynamic high latency WAN as shown in figure \ref{fig:0.1}.

\subsection{Results and Discussion}

\begin{figure}[h]
	\centering
	\includegraphics[scale=0.5]{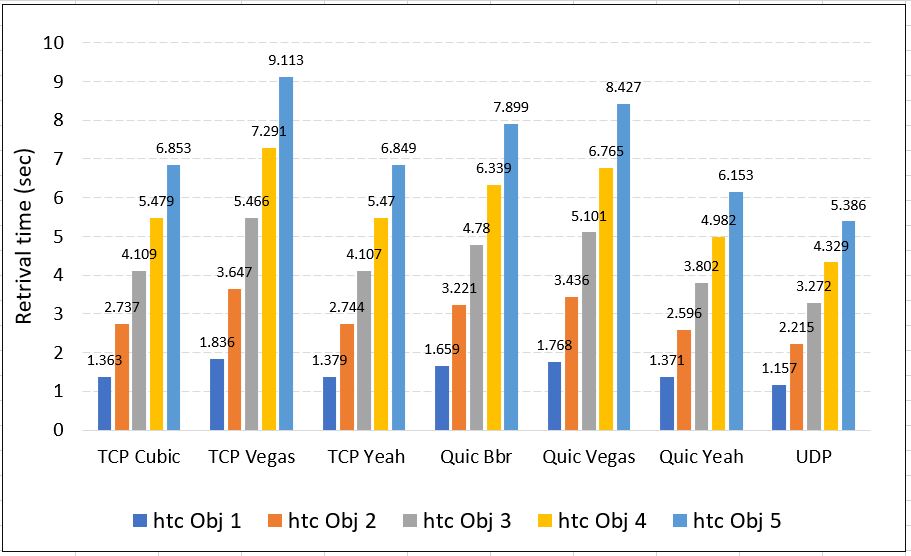}
	\caption{Retrieval time of HTC object by different transport protocols and congestion variants}
	\label{fig:0}
\end{figure}

\begin{figure}[t]
	\centering
	\begin{subfigure}[b]{0.241\textwidth}
		\centering
		\includegraphics[width=\textwidth]{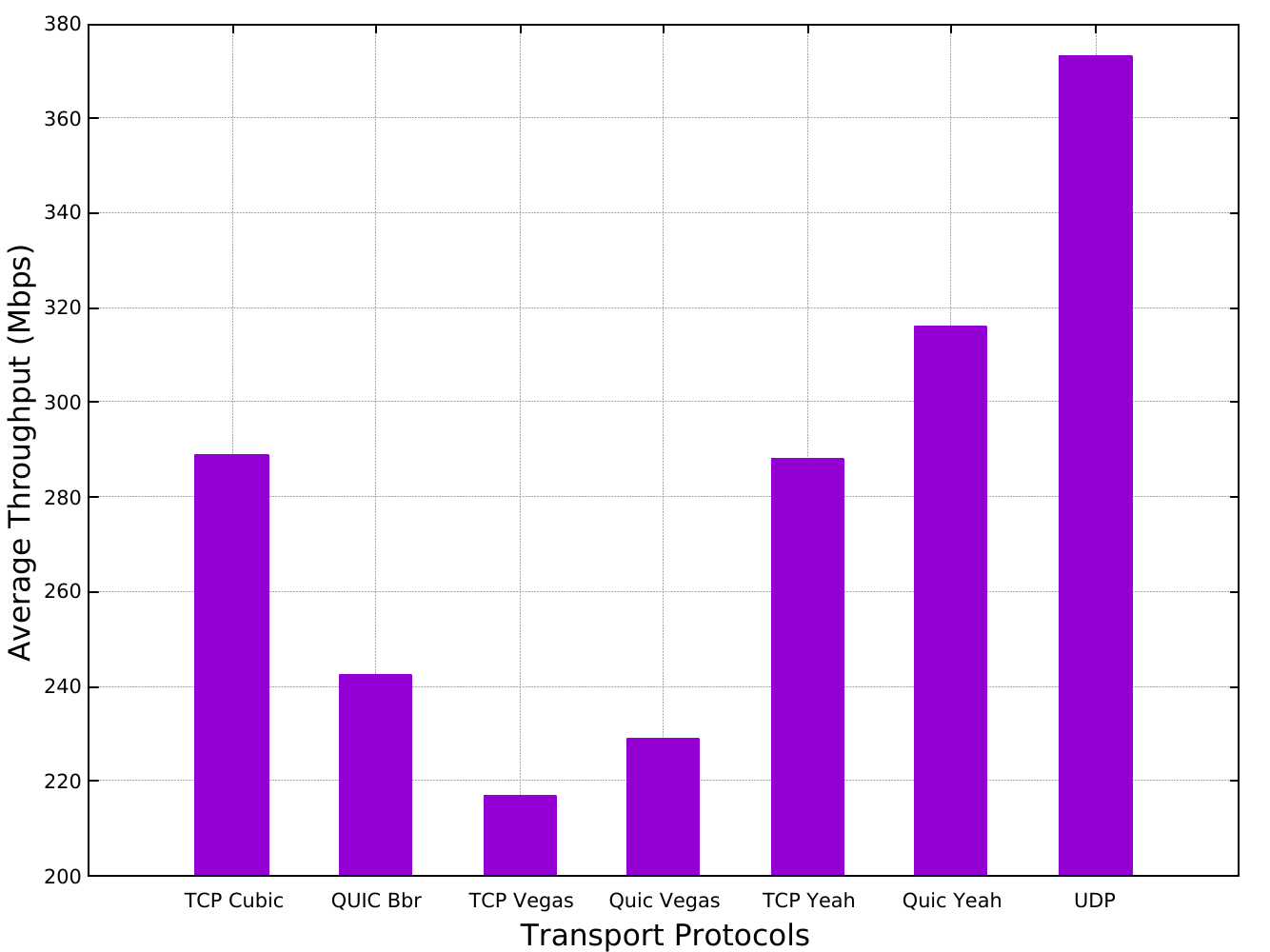}
		\caption{Average throughput in LAN scenario for different congestion variants}
		\label{fig:1of1}
	\end{subfigure}
	\hfill
	\begin{subfigure}[b]{0.241\textwidth}
		\centering
		\includegraphics[width=\textwidth]{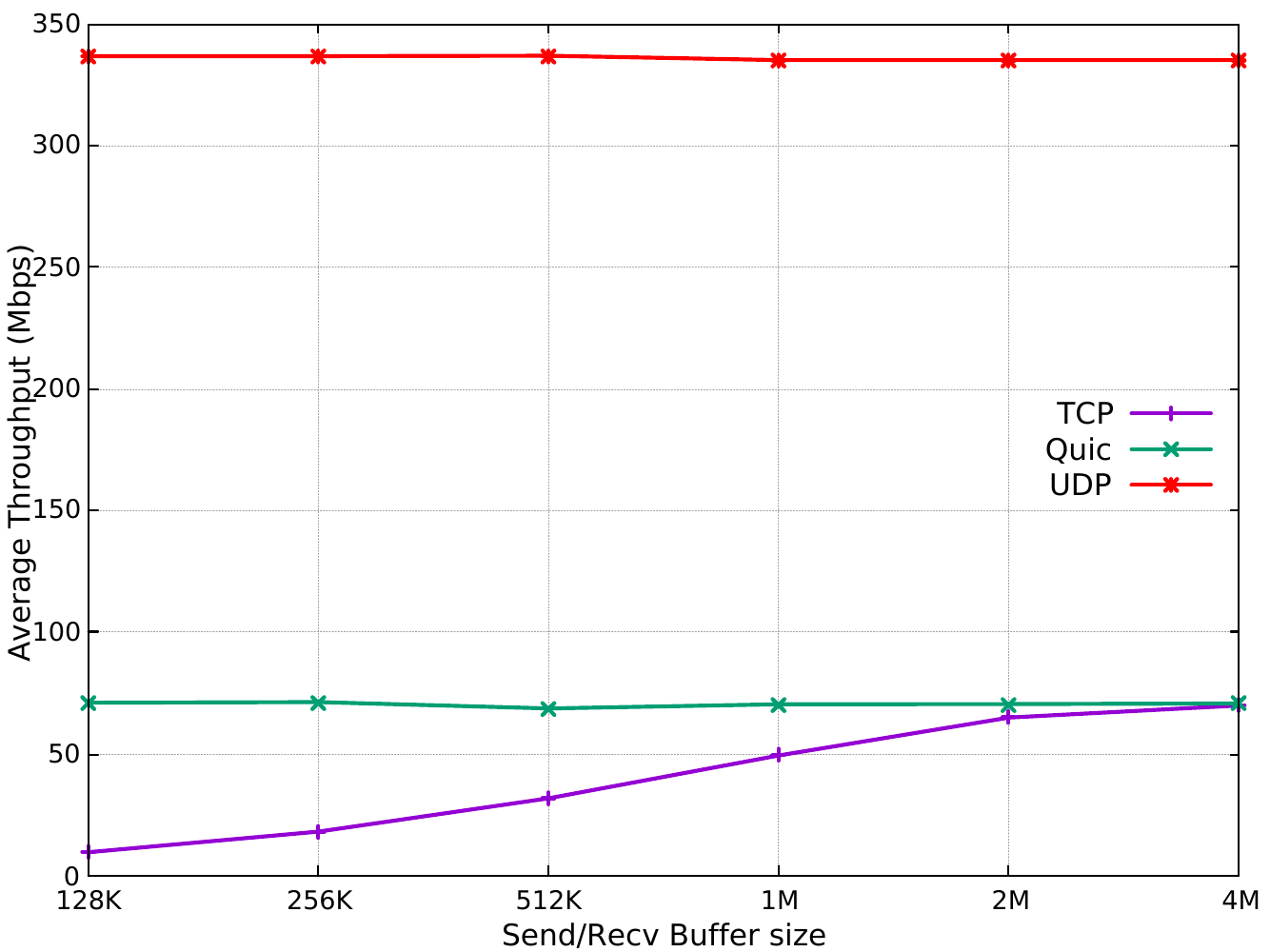}
		\caption{Average throughput in WAN scenario w.r.t send/recv buffer size}
		\label{fig:2of1}
	\end{subfigure}
\caption{Average throughput of transport protocols}
\label{fig:1}
\end{figure}

We are particularly interested in the performance of TCP and QUIC as being the widely used protocols by applications that require guaranteed delivery of data. We use Vegas and YeAH as congestion variants in TCP and QUIC for a fair comparison. We first recorded the retrieval time of the HTC object, which is $400$ Mbits each by UDP, TCP, and QUIC under different congestion variants. The results in figure \ref{fig:0} show that UDP takes less time to retrieve the HTC object as it does not have any flow and congestion control. However, UDP can not be adopted by applications that require guaranteed delivery since UDP encounters huge packet loss and high end-to-end delay, especially when the network is congested. Among TCP and QUIC, QUIC-YeAh and Vegas outperform TCP-YeAH and Vegas, respectively. This is because of the multi-streaming features of the QUIC and the per-stream buffer at the sender and receiver. We also look into the throughput achieved by these protocols in figure \ref{fig:1}. As expected, UDP achieves the highest throughput, $374 mbps$, near the link capacity ($400 mbps$). However, UDP also encounters packet losses, which is not suitable for most immersive applications. QUIC-YeAH achieves $316$ Mbps of throughput compared to $288$ Mbps achieved by TCP-YeAH, as shown in figure \ref{fig:1of1}. Similarly, QUIC-Vegas achieves higher throughput than TCP-vegas though the number is lower than what is achieved by YeAH. We also compared the throughput performance in the WAN scenario as shown in figure \ref{fig:2of1}. The average throughput achieved with respect to varying send/receive buffer size is shown in figure \ref{fig:2of1}. The throughput drops heavily in connection-oriented protocols compared to unreliable connection-less UDP. This heavy drop is because of the high WAN latency and the flow control measures of the connection-oriented protocols. A critical observation from this figure is that unlike TCP, throughput achieved by QUIC is not affected by send/receive buffer size; this is because QUIC maintains per stream buffer, which avoids memory contention at send/receive buffer. Also, we see that the throughput in TCP increases with send/receive buffer size, and the line becomes straight after the buffer size is more than $2$ megabytes. This also attests to our previous observation that at smaller buffer sizes, the throughput decreases because of the memory contention at the buffer that increases the RTT, which activates the flow control to avoid congestion.  

\begin{figure}[t]
	\centering
	\begin{subfigure}[b]{0.241\textwidth}
		\centering
		\includegraphics[width=\textwidth]{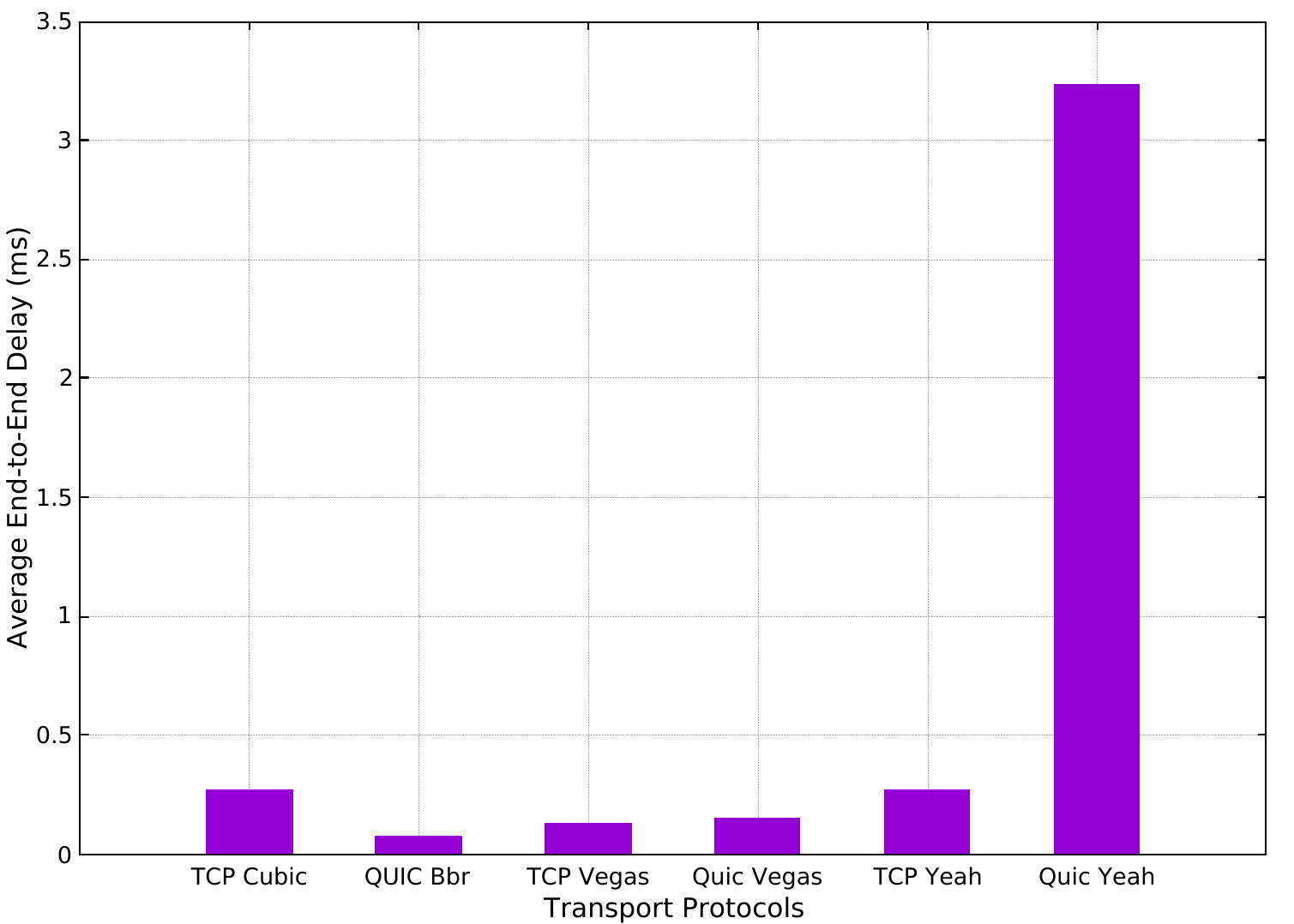}
		\caption{Average end-to-end delay in LAN scenario for different congestion variants}
		\label{fig:1of2}
	\end{subfigure}
	\hfill
	\begin{subfigure}[b]{0.241\textwidth}
		\centering
		\includegraphics[width=\textwidth]{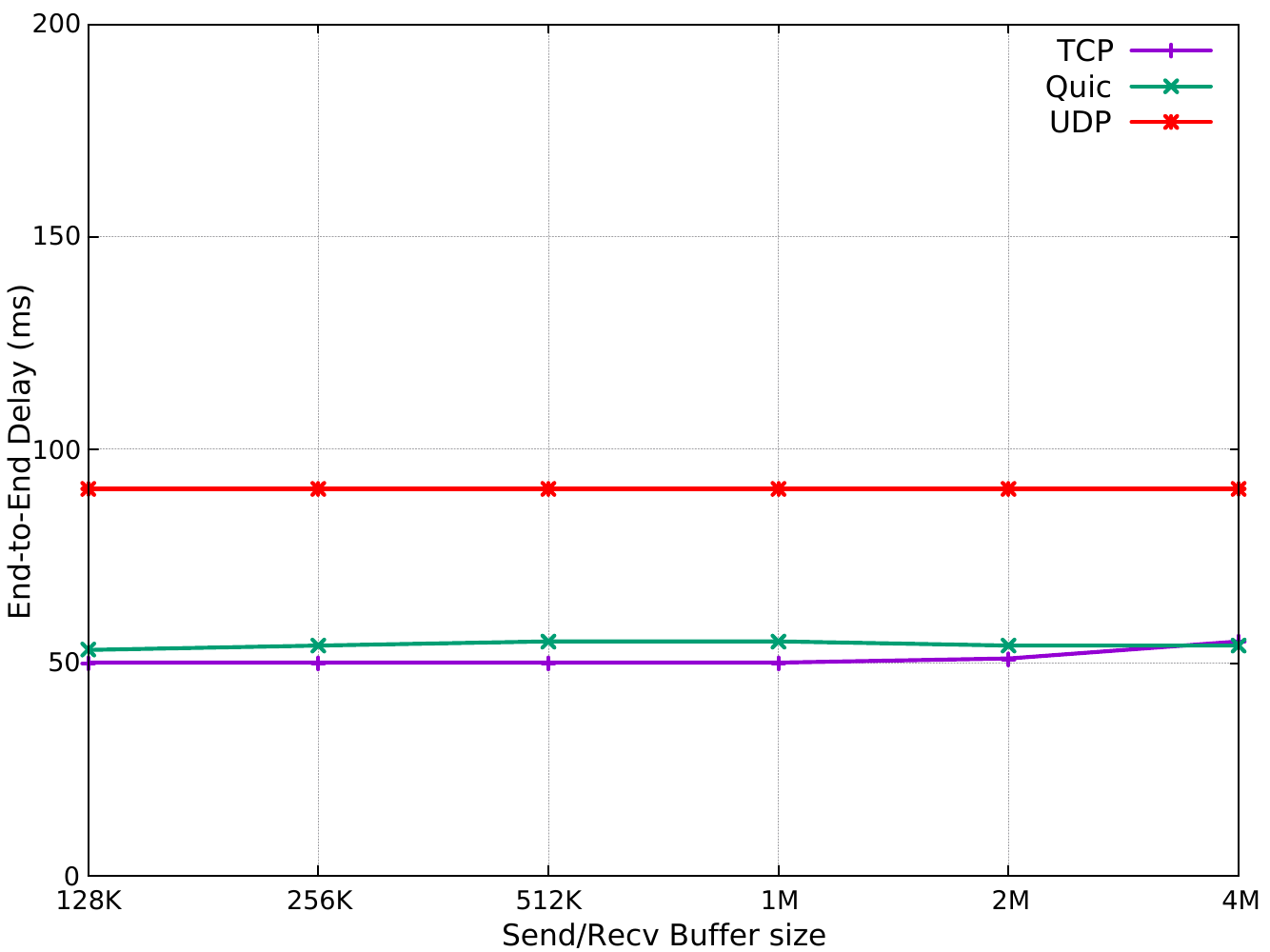}
		\caption{Average end-to-end delay in WAN scenario w.r.t send/recv buffer size}
		\label{fig:2of2}
	\end{subfigure}
	\caption{Average end-to-end delay in transport protocols}
	\label{fig:2}
\end{figure}

\begin{figure}[t]
	\centering
	\begin{subfigure}[b]{0.241\textwidth}
		\centering
		\includegraphics[width=\textwidth]{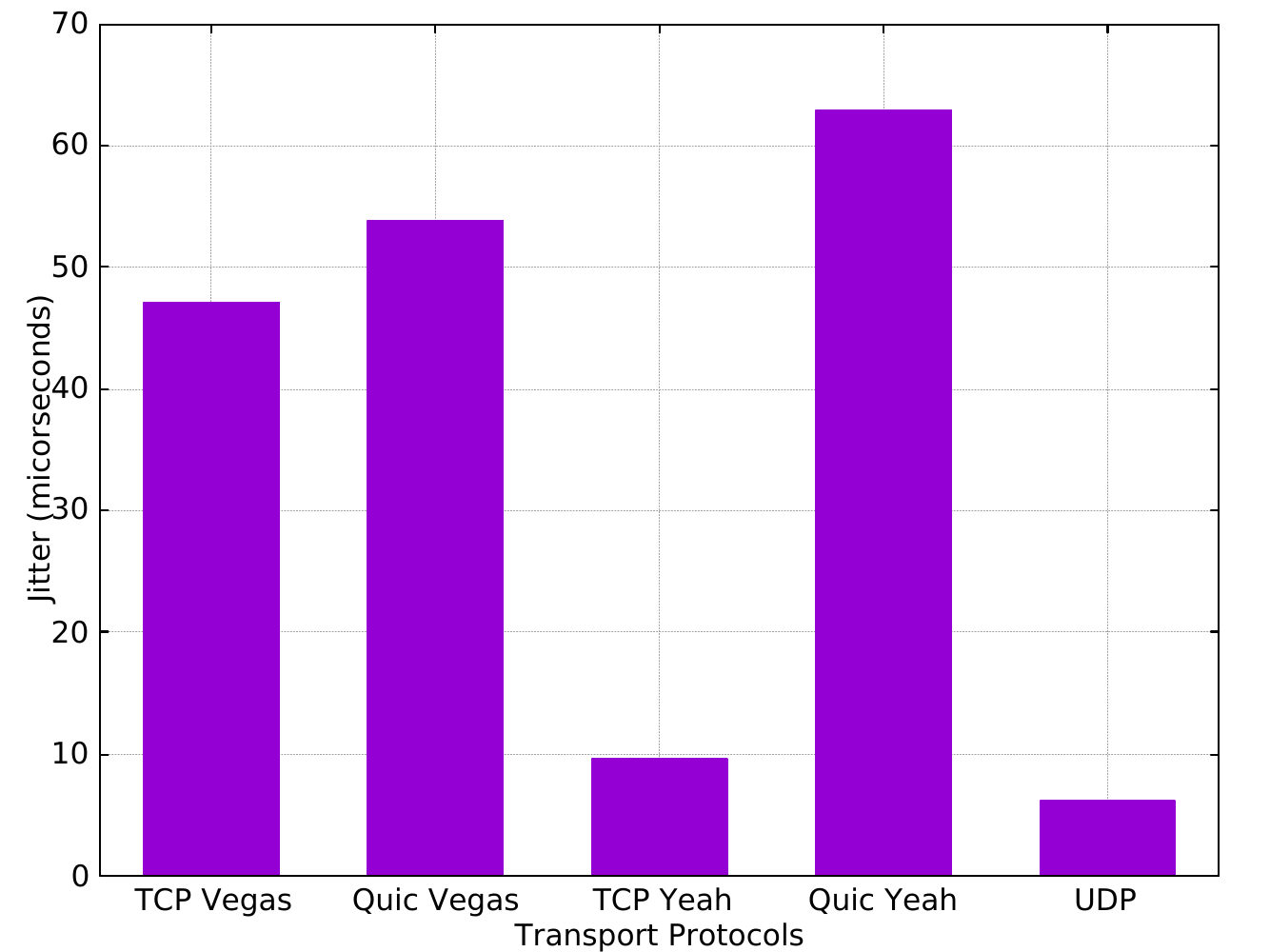}
		\caption{Average jitter in LAN scenario for different congestion variants}
		\label{fig:1of3}
	\end{subfigure}
	\hfill
	\begin{subfigure}[b]{0.241\textwidth}
		\centering
		\includegraphics[width=\textwidth]{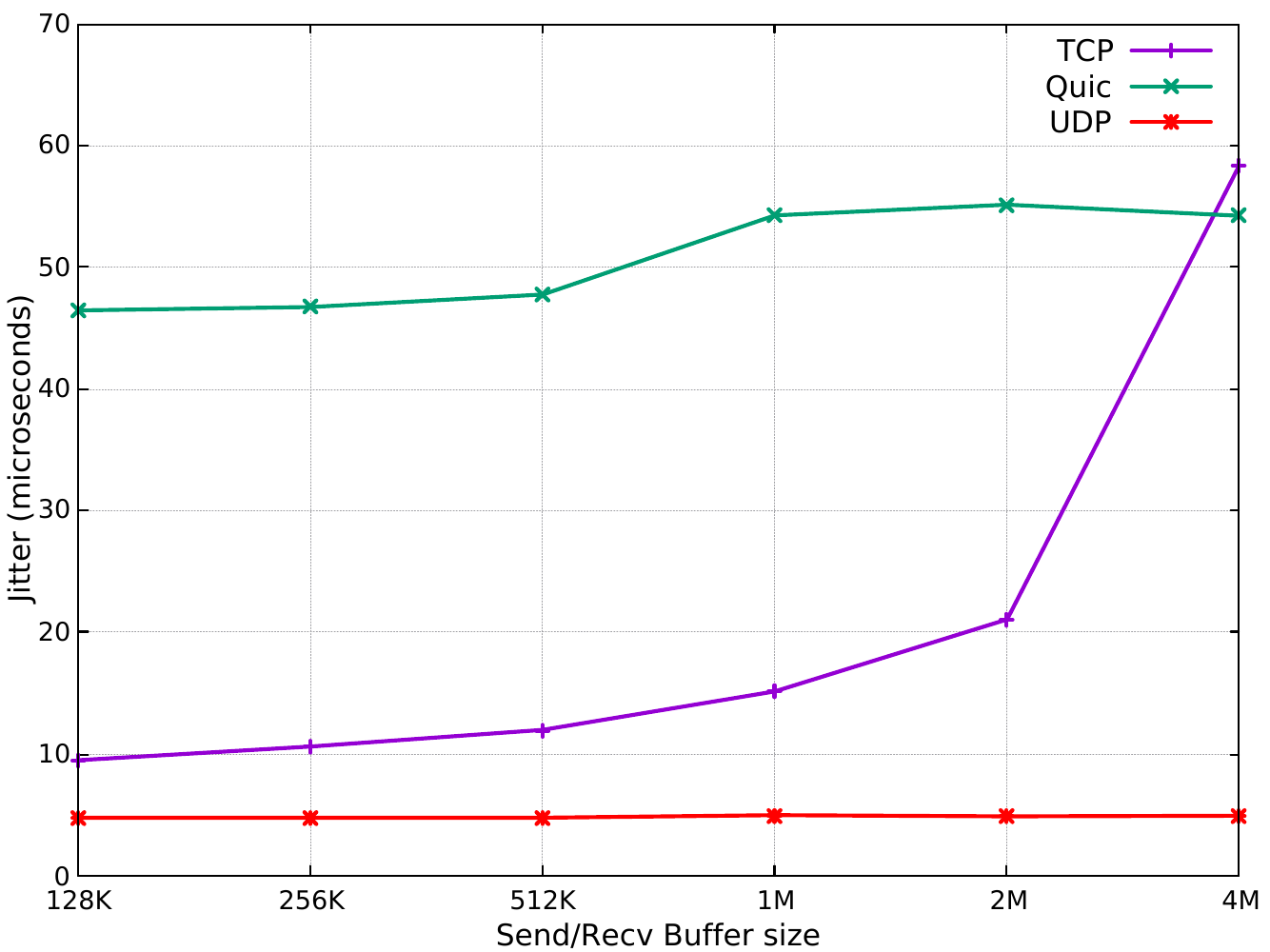}
		\caption{Average jitter in WAN scenario w.r.t send/recv buffer size}
		\label{fig:2of3}
	\end{subfigure}
	\caption{Average jitter in transport protocols}
	\label{fig:3}
\end{figure}

\begin{figure}[t]
	\centering
	\begin{subfigure}[b]{0.241\textwidth}
		\centering
		\includegraphics[width=\textwidth]{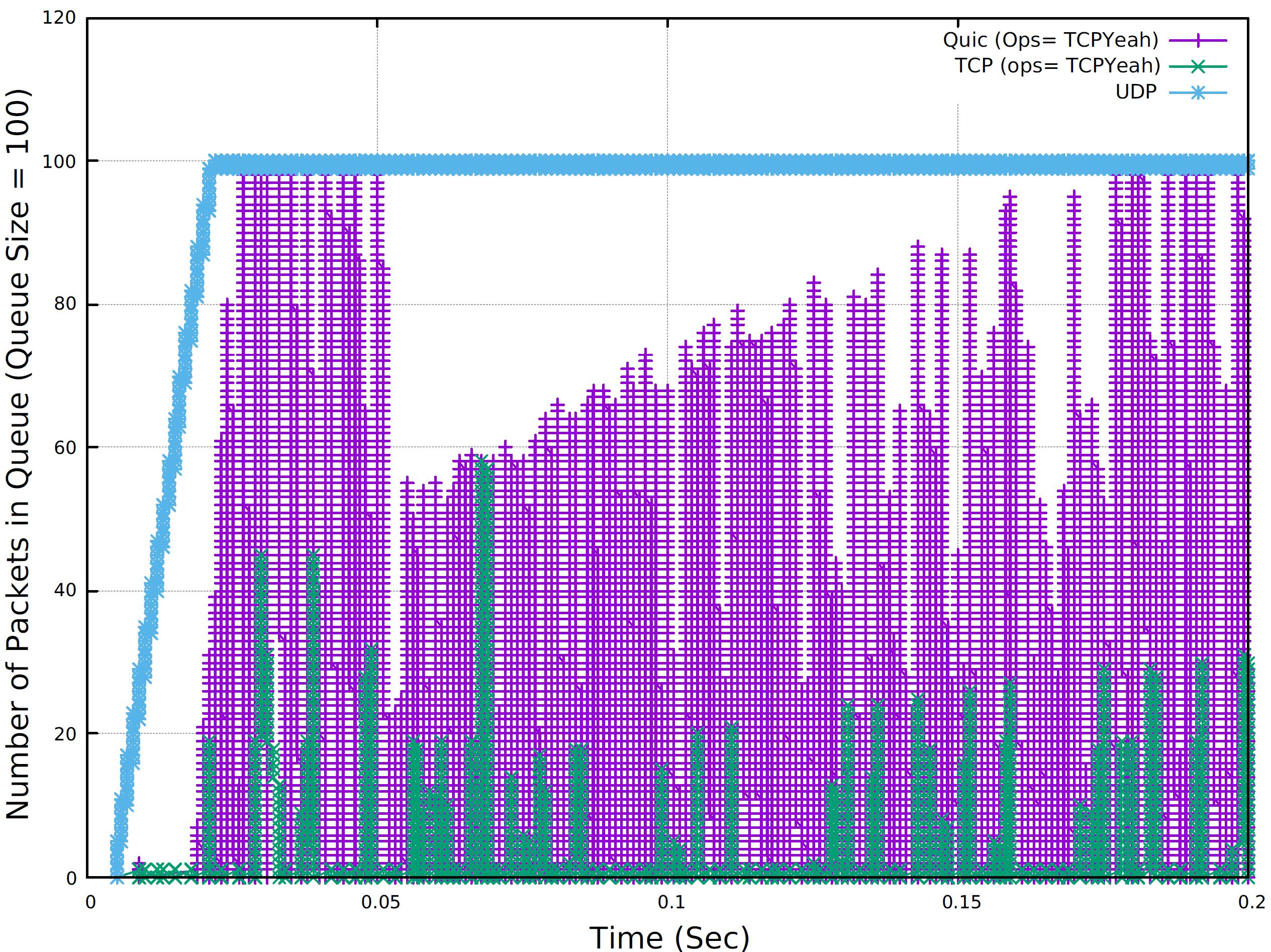}
		\caption{Queue size change in LAN scenario for transport protocols}
		\label{fig:1of4}
	\end{subfigure}
	\hfill
	\begin{subfigure}[b]{0.241\textwidth}
		\centering
		\includegraphics[width=\textwidth]{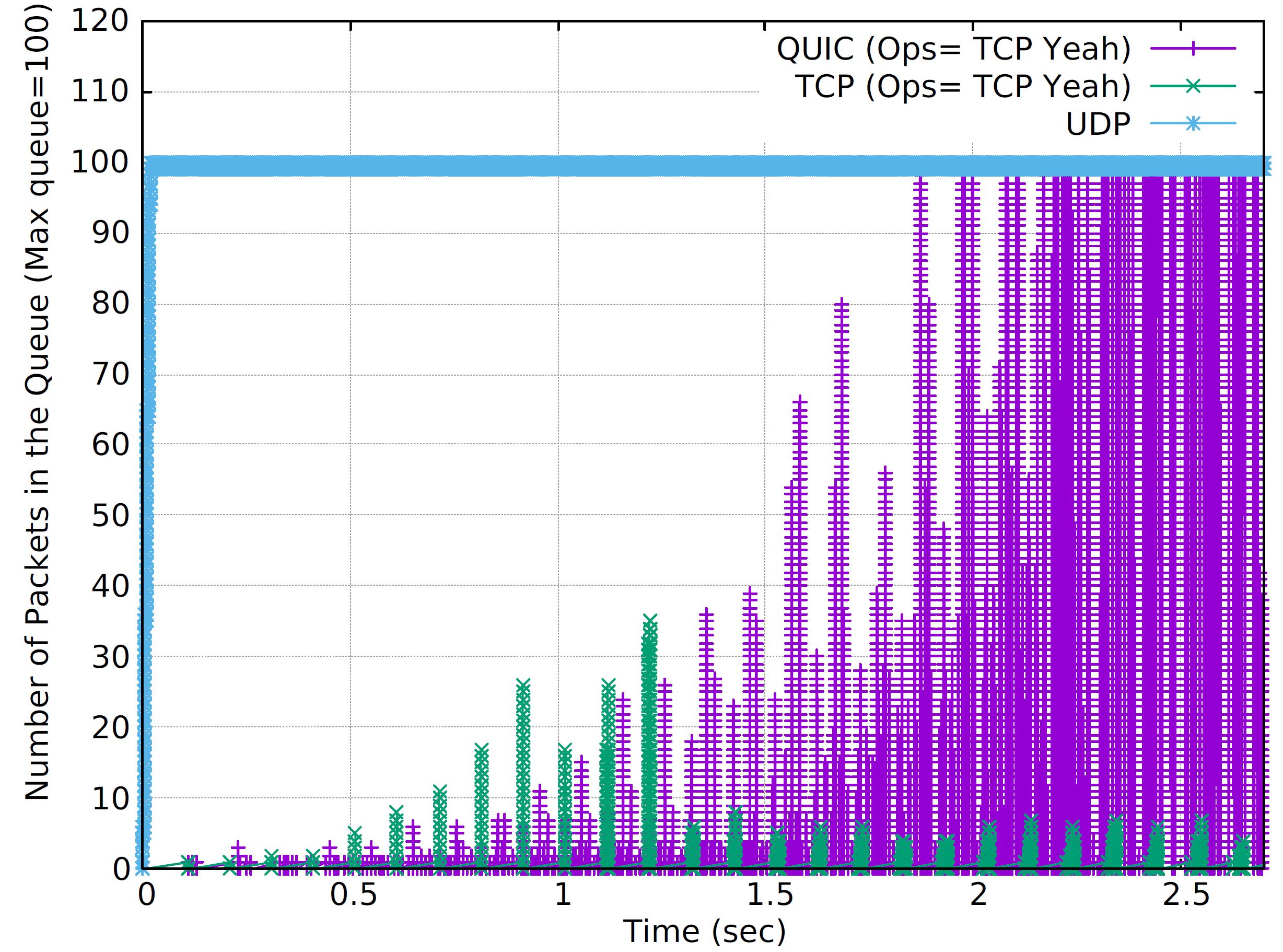}
		\caption{Queue size change in WAN scenario for transport protocols}
		\label{fig:2of4}
	\end{subfigure}
	\caption{Queue size change in transport protocols (Congestion option is "Yeah")}
	\label{fig:4}
\end{figure}

\begin{figure}[t]
	\centering
	\begin{subfigure}[b]{0.241\textwidth}
		\centering
		\includegraphics[width=\textwidth]{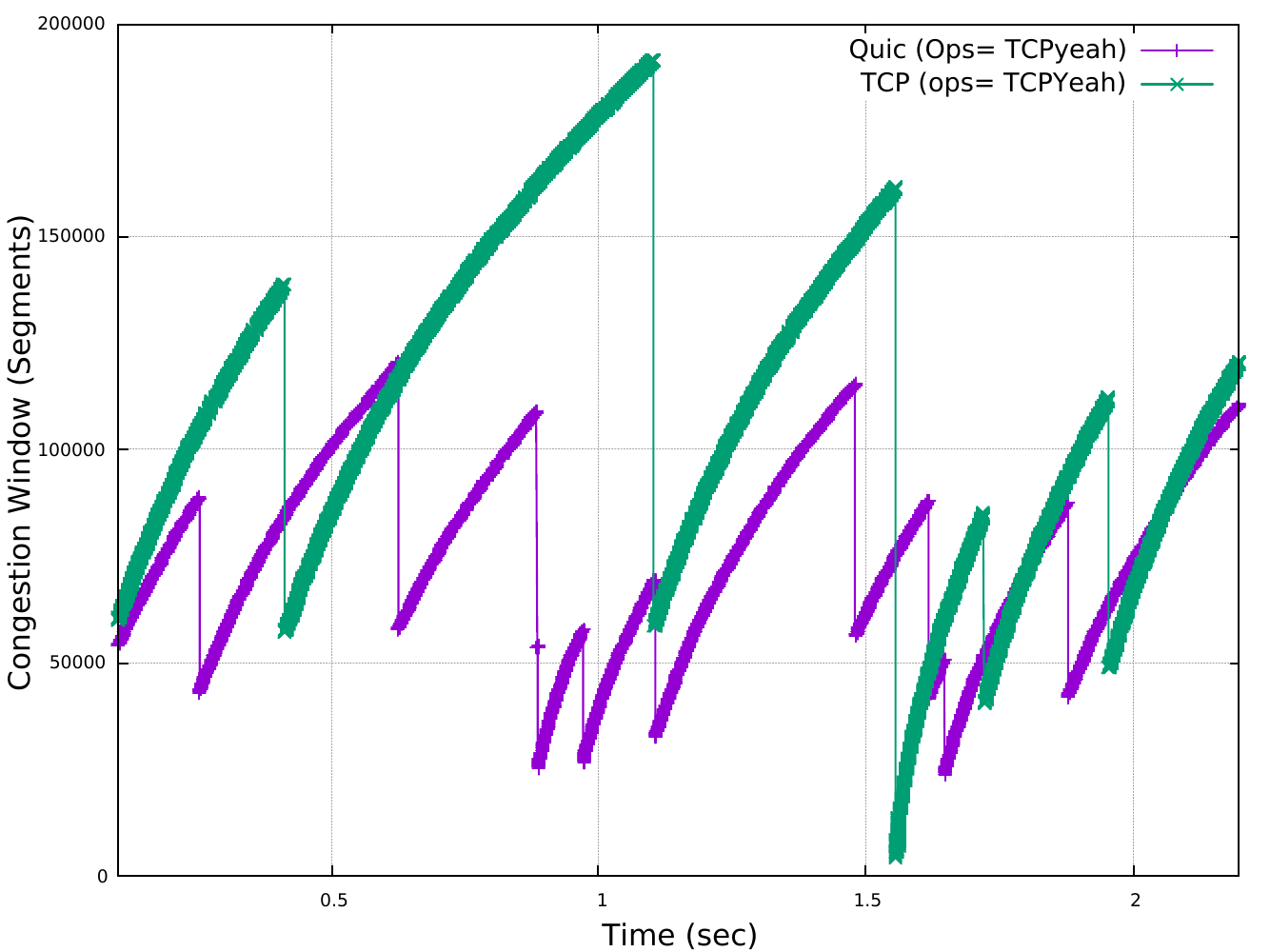}
		\caption{Congestion window in LAN scenario for transport protocols}
		\label{fig:1of5}
	\end{subfigure}
	\hfill
	\begin{subfigure}[b]{0.241\textwidth}
		\centering
		\includegraphics[width=\textwidth]{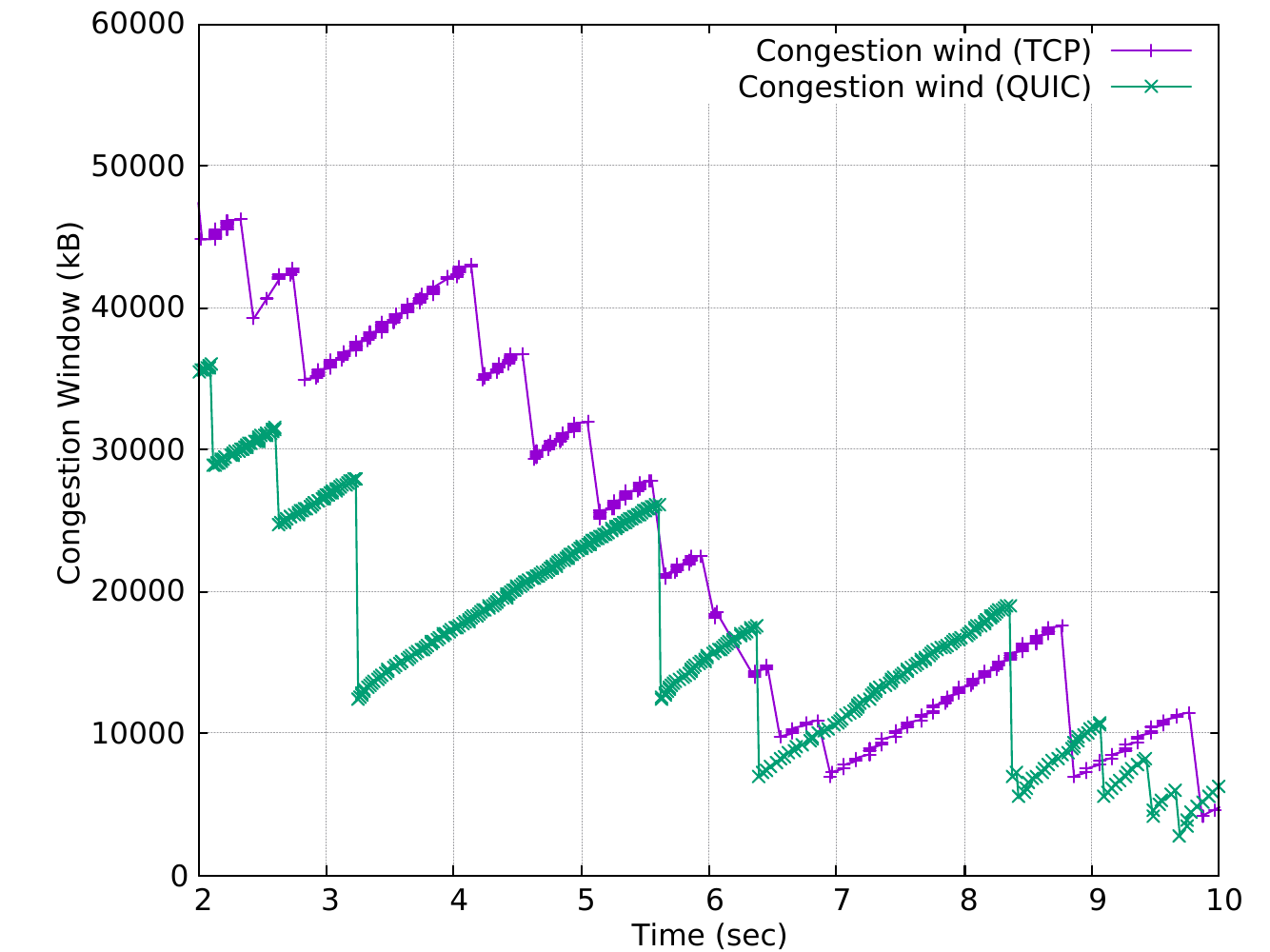}
		\caption{Congestion window in WAN scenario for transport protocols}
		\label{fig:2of5}
	\end{subfigure}
	\caption{Congestion window in transport protocols (Congestion option is "Yeah")}
	\label{fig:5}
\end{figure}

We now investigate further into the protocols by analyzing the end-to-end delay. In LAN scenario figure \ref{fig:1of2}, QUIC-YeAH has a high end-to-end delay compared to TCP-YeAH. We investigated the reasons behind this and concluded that this high end-to-end delay is because QUIC-YeAH keeps the pipe full to achieve high throughput, which causes queue latency at nodes that add up to the average end-to-end delay. Figure \ref{fig:4} shows queue occupancy. In figure \ref{fig:4}, UDP keeps the queue full all the time, which puts the network in a congestion state and causes packet loss and very high end-to-end delay ($247 ms$). The queue size in the experiment is $100$ packets. QUIC-YeAH keeps the queue nearly full compared to TCP-YeAH. Once the queue becomes full, QUIC flow control slows down the ingestion of packets to avoid congestion. The reasons for the average end-to-end delay trend of the protocols become more clear from the results of queue occupancy. This behavior of QUIC also affects the average jitter, as shown in figure \ref{fig:3}. The jitter in QUIC is high compared to TCP as QUIC keeps the queue near its full capacity and then abruptly controls the flow to avoid congestion, unlike TCP, which keeps the queue capacity at nearly half. This causes variation in packet arrival at the receiver. On the other hand, the jitter in UDP is very low for obvious reasons. The average jitter results in the WAN scenario figure \ref{fig:2of3} are rather more interesting. Our observations from the figure are that jitter increases with the increase in throughput; as we can see from the throughput results (figure \ref{fig:2of1}), QUIC achieves high throughput than TCP when the send/receive buffer size varies from $128$K to $2$M. A similar trend can be seen in average jitter. QUIC faces high jitter than TCP when the send/receive buffer size varies from $128$K to $2$M (figure \ref{fig:2of3}). At $4$M send/receive buffer size, the average throughput of TCP equates to QUIC throughput, and the average jitter of TCP shows an abrupt increase and equates to QUIC's average jitter. On further investigation, this pattern is caused by the similar queue occupancy of TCP and QUIC at $4$ MB of send/receive buffer size.

\begin{figure}[t]
	\centering
	\includegraphics[scale=0.4]{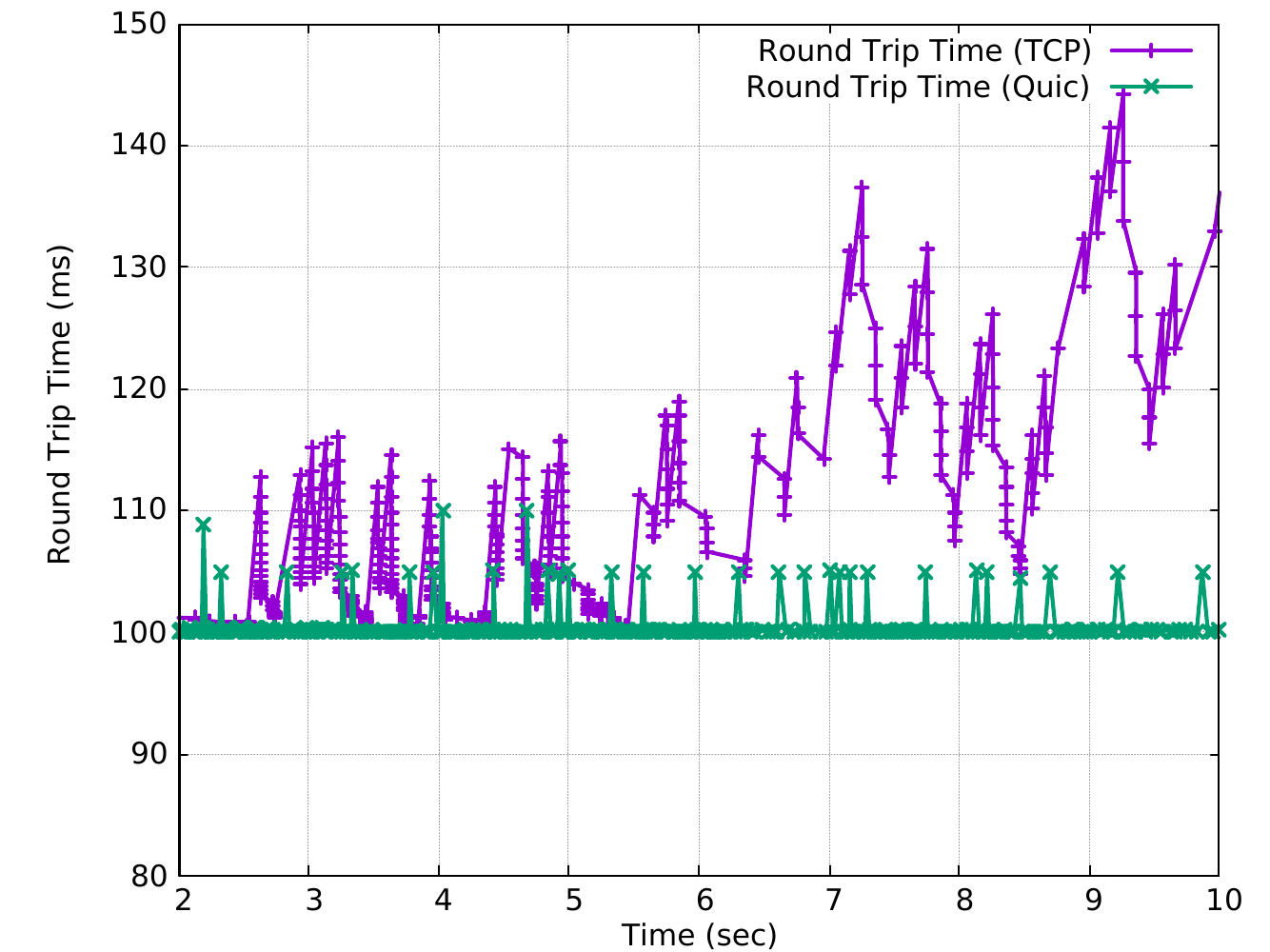}
	\caption{Round trip time in WAN scenario (Congestion option is "Yeah")}
	\label{fig:6}
\end{figure}

\begin{figure}[t]
	\centering
	\begin{subfigure}[b]{0.241\textwidth}
		\centering
		\includegraphics[width=\textwidth]{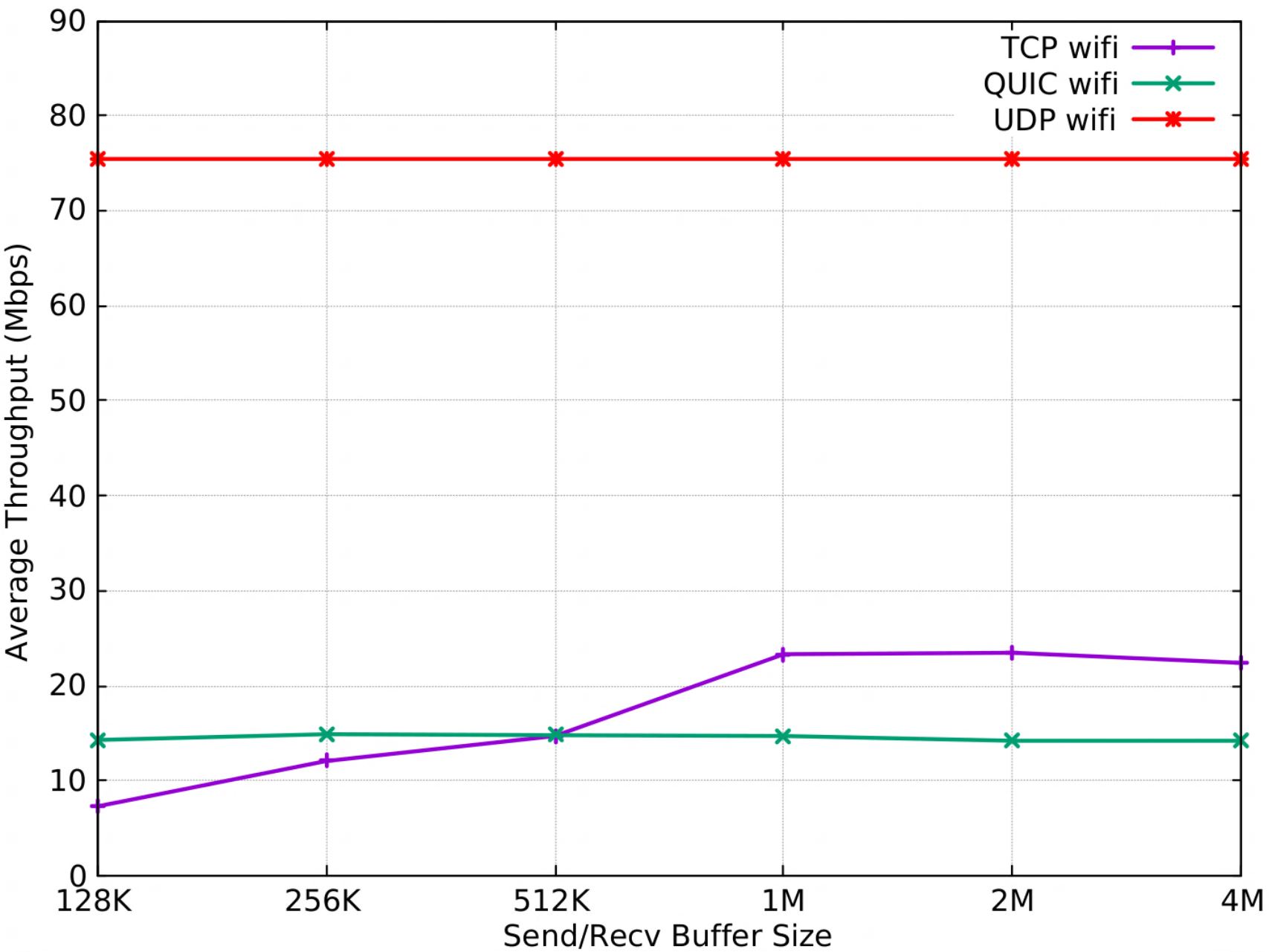}
		\caption{Average throughput for wireless client}
		\label{fig:1of7}
	\end{subfigure}
	\hfill
	\begin{subfigure}[b]{0.241\textwidth}
		\centering
		\includegraphics[width=\textwidth]{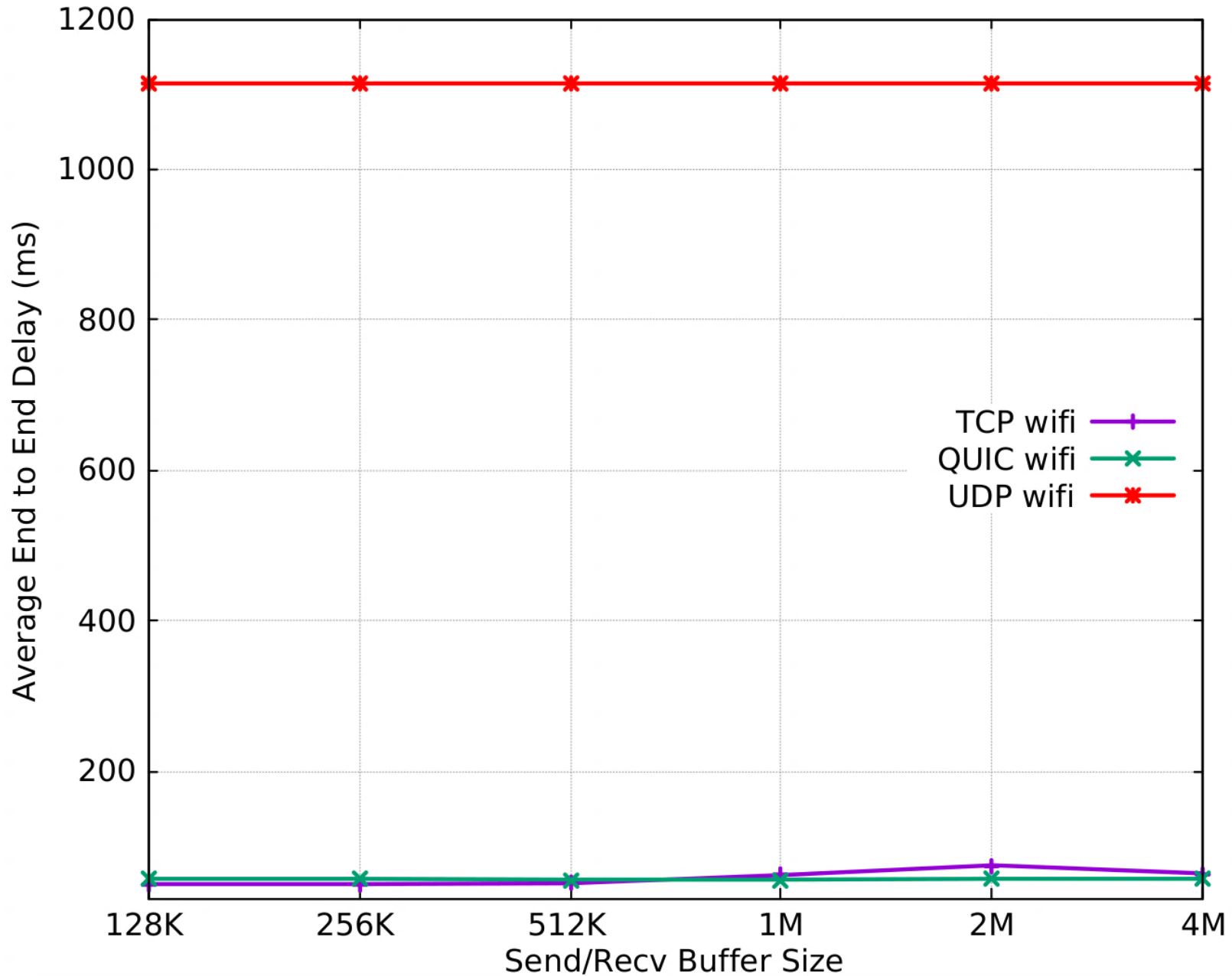}
		\caption{End-to-end delay for wireless client}
		\label{fig:2of7}
	\end{subfigure}

	\caption{Throughput and delay for wireless client}
	\label{fig:7}
\end{figure}

\begin{figure}[t]
	\centering
	\begin{subfigure}[b]{0.241\textwidth}
		\centering
		\includegraphics[width=\textwidth]{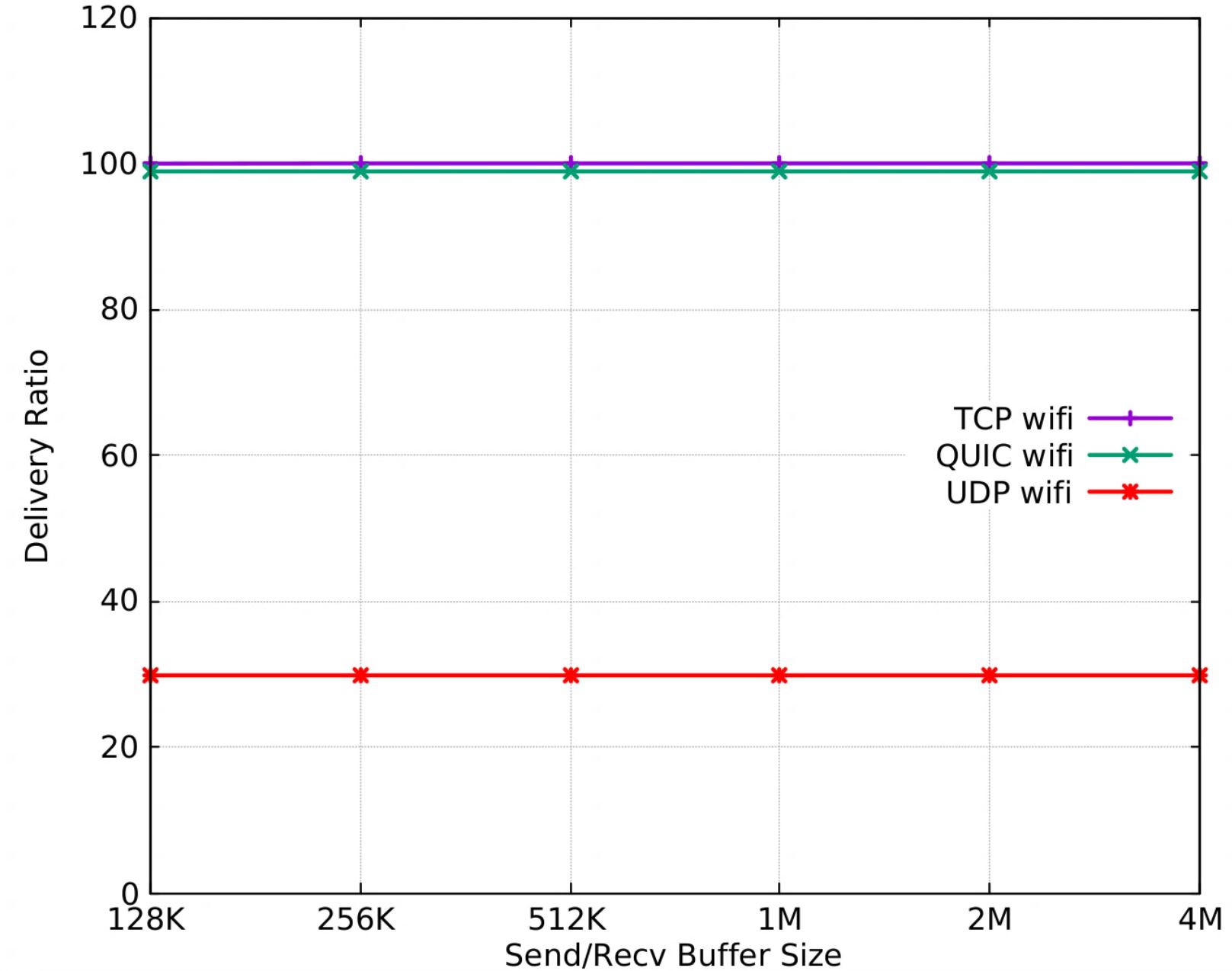}
		\caption{Delivery ratio for wireless client}
		\label{fig:1of8}
	\end{subfigure}
	\hfill
	\begin{subfigure}[b]{0.241\textwidth}
		\centering
		\includegraphics[width=\textwidth]{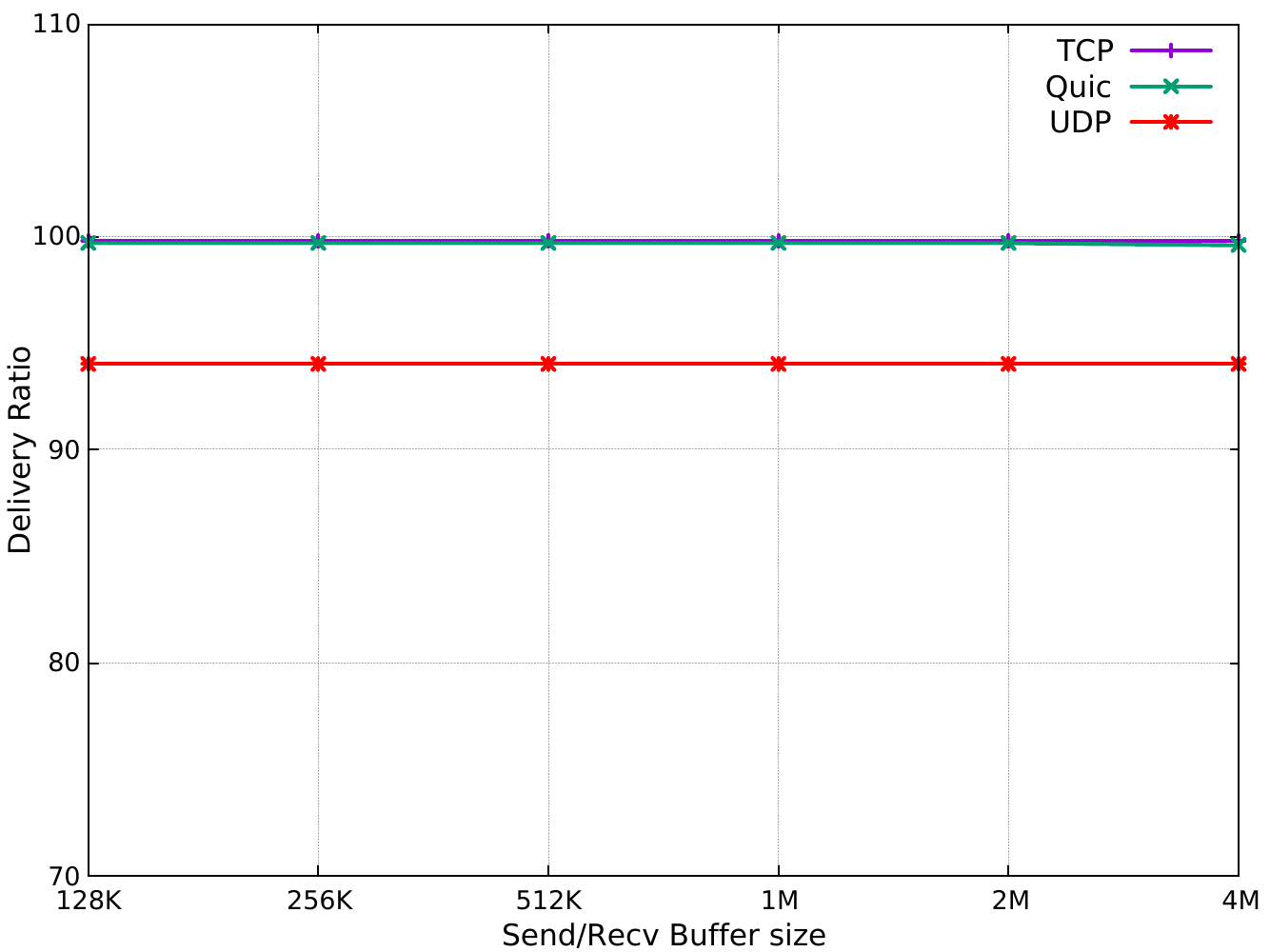}
		\caption{Delivery ratio for wired client}
		\label{fig:2of8}
	\end{subfigure}
	
	\caption{Delivery ratio for wireless and wired client}
	\label{fig:8}
\end{figure}

We also show the congestion window change in figure \ref{fig:5} to make it easy to understand the congestion control behavior of YeAH. The congestion window increases exponentially and steeply drops when congestion is detected. We further show the round trip time (RTT) variation in QUIC and TCP in figure \ref{fig:6}. The RTT in TCP is a little higher than QUIC because of TCP's single send/receive buffer compared to QUIC, which has a per-stream buffer. A packet loss in TCP stalls other packets received correctly, causing an increase in RTT. On the other hand, a packet loss in one stream of QUIC does not affect packet reception in other streams, which is why QUIC has stable RTT, as shown in the figure below. Connection-oriented reliable transport protocols perform worst in the wireless environment due to heavy losses of packets at the wireless channel. Figure \ref{fig:7} shows the average throughput in a wireless environment where only the client is connected through WiFi. The throughput of all three protocols drops heavily because of the losses due to the wireless interface. UDP throughput, which was $340$ Mbps, drops to $76$ Mbps when retrieving data using WiFi. Similarly, QUIC and TCP throughput drops heavily, as shown in figure \ref{fig:1of7}. The trend is similar to that in the wired environment, except that TCP at a larger send/receive buffer size achieves slightly higher throughput. The end-to-end Delay of UDP is very high because of the frequent losses at the wireless interface and re-transmission at the mac layer. The end-to-end delay of QUIC and TCP shows a similar trend as in a wired environment; however, the average end-to-end delay increases by approximately $15$ ms, which is the effect of losses at wireless interfaces and latency added by mac layer re-transmission. Finally, figure \ref{fig:8} shows the Delivery ratio of wired and wireless clients in the WAN scenario. Since TCP and QUIC guarantee packet delivery gives a $100$\% delivery ratio at the cost of higher end-to-end delay. On the other hand, the delivery ratio of UDP is worst in the wireless environment, as shown in figure \ref{fig:1of8}. The delivery ratio of the wired client is shown in figure \ref{fig:2of8}, which is as expected $100$\% for TCP and QUIC while $94$\% for UDP, keeping in mind that the Linux traffic control layer was enabled, resulting in a high delivery ratio for UDP.

\section{Roadmap for Designing Novel Transport Protocols for Immersive Applications}

Network applications and their requirements have changed rapidly in the past decade resulting in new challenges for the existing transport protocols. Soon most of the Internet traffic will be generated by immersive applications, including holographic-type communication. Therefore, a new high-performance transport design is needed from scratch to tackle the challenges of modern immersive applications and their possible future demands.

Congestion control algorithms are the key to transport protocols. However, their performance is affected by erroneous congestion detection. For example, RTT variation caused by MAC layer re-transmission and packet loss caused by a lossy wireless link sometimes triggers the congestion control algorithm. Artificial intelligence may be a valuable alternative to detect congestion in the network. Similarly, out-of-order packet delivery triggers false congestion. A new robust mechanism for congestion detection and out-of-order packet delivery must be a feature of the transport protocols. The congestion control algorithm must consider the challenges caused by the wireless environment. 
The fairness concept in TCP must be revisited in the design of new transport protocols. A significant amount of Internet traffic is caused by software updates and some other mice flow, which are highly delay tolerant and can be assigned fewer network resources than other connections. Furthermore, in the case of multi-streaming, per stream buffer can further enhance the performance. 

Holographic objects can be viewed from different angles, and not all areas of the HTC object are visible to the user based on the user's viewing position. Transport protocols should be able to adapt themselves to decide which part of the content to stream first and at what quality. However, this requires the protocols to predict the user's viewpoint, and movement \cite{b12}. A few essential features of high-performance transport protocols are discussed below.

\subsection{Application-specific Requirements }
\begin{figure}[t]
	\centering
	\includegraphics[scale=0.4]{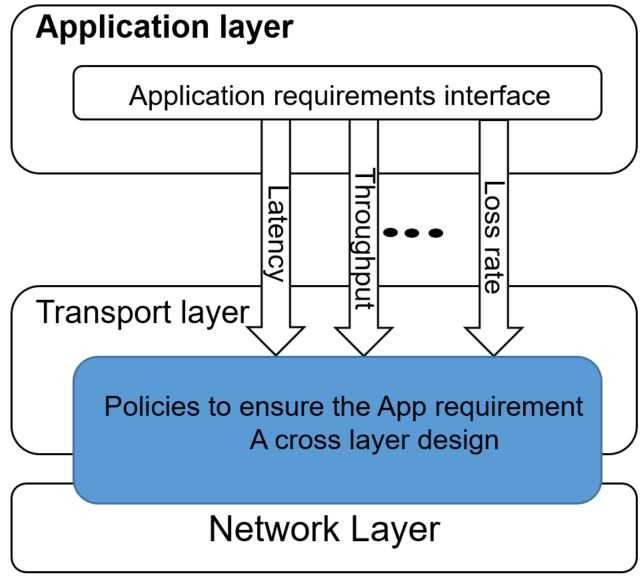}
	\caption{Application specific requirements}
	\label{fig:9}
\end{figure}

Different network applications may have different requirements, and fixed rules of protocols will not be able to do justice to the application requirements. For example, some applications may require high reliability but may not require ultra-low latency. Some applications may be loss tolerant but may require high bandwidth and low latency, such as online games and OTT services. Similarly, some applications, such as telesurgery, may require ultra-low latency and strict deadlines. Therefore, new high-performance transport protocols must consider application requirements in their design and try to meet the application requirements. A viable solution is a cross-layer design to meet the application-specific requirements, as shown in figure \ref{fig:9}. The network layer can provide explicit knowledge of the network condition, such as congestion state, open paths, multi-path etc. The transport layer can easily exploit this knowledge to ensure application-specific requirements.

\subsection{Multicast support} 
\begin{figure}[t]
	\centering
	\begin{subfigure}[b]{0.241\textwidth}
		\centering
		\includegraphics[width=\textwidth]{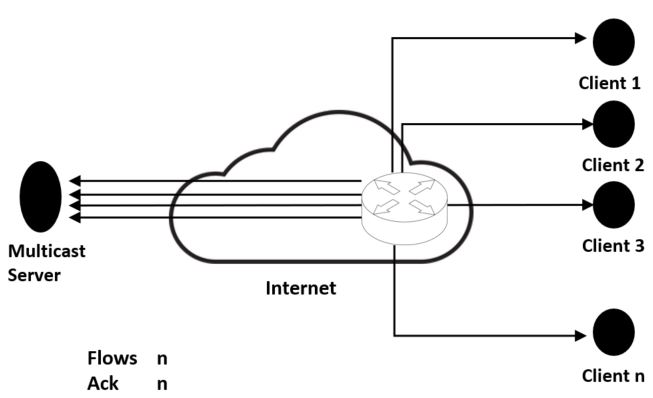}
		\caption{No multicast support}
		\label{fig:1of10}
	\end{subfigure}
	\hfill
	\begin{subfigure}[b]{0.241\textwidth}
		\centering
		\includegraphics[width=\textwidth]{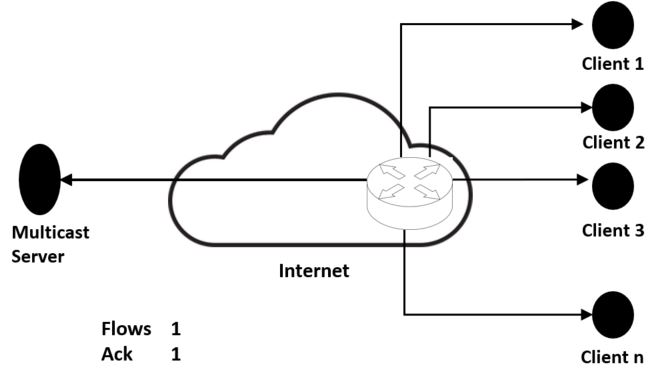}
		\caption{Multicast support}
		\label{fig:2of10}
	\end{subfigure}
	\caption{Multicast support}
	\label{fig:10}
\end{figure}

Most connection-oriented reliable protocols do not support multicast communication. In traditional transport protocols, if there is one sender and $n$ receivers, a single packet will trigger $n$ acknowledgment packets and there will be $n$ total flows as shown in figure \ref{fig:1of10}. The design of new high-performance transport protocols must include multicast support as a huge percentage of communication on the Internet is based on multicast. Multicast saves bandwidth as the server sends only one content packet, which is multicasted by the network to a multicast group as shown in figure \ref{fig:2of10}. A new mechanism of ack and retransmission of lost packets should be implemented. For example, the gateway router can receive individual acks and sends one ack back to the server. Moreover, a small buffer could be used at the multicast edge router to retransmit the lost packet. This will save massive bandwidth. 

\subsection{Segment level reliability}
The existing transport protocols transmit data without knowing its importance in a connection. For example, some content may go unnoticed by the users and may be sent at lower quality or even be ignored if that particular segment is lost. A considerable amount of bandwidth can be saved if the protocol deals with these less critical segments differently. This feature will impact the performance of transport protocols for immersive applications as a considerable amount of volumetric data can be sent at lower quality and can be ignored if a loss occurs, saving considerable bandwidth.

\subsection{Multi-streaming and connection migration}
\begin{figure}[t]
	\centering
	\includegraphics[scale=0.45]{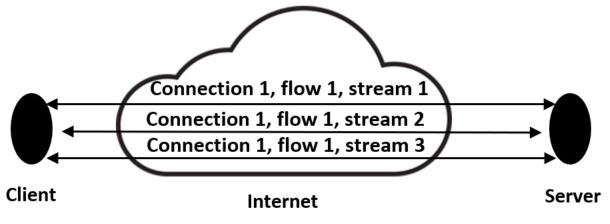}
	\caption{Multi streaming in transport protocol}
	\label{fig:11}
\end{figure}

\begin{figure}[t]
	\centering
	\includegraphics[scale=0.45]{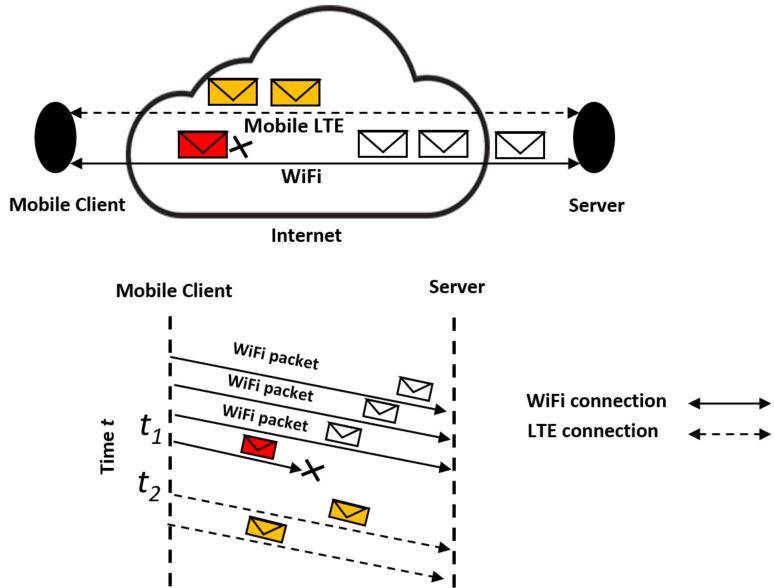}
	\caption{Connection migration in transport protocol}
	\label{fig:12}
\end{figure}

Multi-streaming is sending data of a single flow on multiple streams, as elaborated in figure \ref{fig:11}. This not only achieves high throughput but also avoid HoL blocking issue. Moreover, high-priority data in a flow could also benefit from multi-streaming by assigning it to a single high-priority stream. The high-priority stream may have different network requirements than the other streams in the same flow. For example, critical data in a flow which is loss sensitive must be assigned to a specific stream with loss-sensitive requirements from the application. Similarly, connection migration is also a must-needed feature in any new transport protocol because of the widespread of the Internet in the form of WiFi and LTE. With the increasing popularity of mobile devices with multiple Internet connections simultaneously, connection migration from one connection over to a more stable connection is vital for reliability, as explained in figure \ref{fig:12}. Traditional transport protocols do not have multi-streaming and connection migration features. However, the recent release of QUIC protocols has both of the mentioned features. These features are vital for modern high-performance transport protocols for immersive applications. Application-level protocols like HTTP/2 supports multi-streaming to avoid HoL blocking issue; however, these multiple streams are multiplexed on a single TCP stream at the transport layer limiting the benefit of HTTP/2 multi-streaming. Connection migration and keeping recent connection states to avoid handshake delay when the connection times out due to a congested wired link or lossy wireless channel must be included in the design of the transport protocols. 

\section{Conclusion and Future Work}

Holographic-type communication is going to be vital for future hologram-based networked applications. Holograms and extended reality are emerging media types that will replace the current 2D and 3D volumetric content. However, this kind of media requires new network architecture and protocols for guaranteed user experience. In this paper, we have discussed the network requirements of the HTC, then we have provided an in-depth analysis of the standard transport protocols, i.e., UDP, TCP, and QUIC. We compared their performance and provided a critical discussion on the results. Based on the results and discussion, we presented a roadmap for the design of a novel high-performance transport protocol. This detailed study will help the researchers to get a deep insight into the current transport protocols for immersive applications and takeaways for designing a novel, efficient transport protocol.  

In future work, we are going to design and develop a new high-performance transport protocol for immersive applications.

\section*{Acknowledgment}

This work was supported by Electronics and Telecommunications Research Institute
(ETRI) grant funded by ICT R\&D program of MSIT/IITP[2021-0-00715, Development of End-to-End
Ultra-high Precision Network Technologies].

\end{document}